\documentclass[prd,preprint,superscriptaddress,showpacs,byrevtex]{revtex4} 
\usepackage{epsfig} 
\newcommand{\be}{\begin{equation}} 
\newcommand{\ee}{\end{equation}} 
\newcommand{\bea}{\begin{eqnarray}} 
\newcommand{\eea}{\end{eqnarray}} 

\begin{document} 

\begin{flushright}
YITP-SB-05-36
\end{flushright}

\title{\noindent Non-Perturbative Gluon pair production from a
Constant Chromo-Electric Field via the Schwinger Mechanism in Arbitrary Gauge}

\author{Fred Cooper} \email{cooper@santafe.edu} 
\affiliation{ Santa Fe Institute, Santa Fe, NM 87501, USA}

\author{Gouranga C. Nayak} \email{nayak@insti.physics.sunysb.edu} 
\affiliation{ C. N. Yang Institute for Theoretical Physics, Stony Brook University, SUNY, Stony Brook, 
NY 11794-3840, USA } 

\date{\today} 

\begin{abstract} 
We study  the non-perturbative production of gluon pairs  from a constant  SU(3) chromo-electric  background field via the Schwinger mechanism. We fix the covariant background gauge with an arbitrary  gauge parameter $\alpha$.   We determine the transverse momentum 
distribution of the gluons, as well as the total probability of creating pairs per unit space time volume.  We find that the result is independent of the covariant gauge
parameter $\alpha$ used to define arbitrary covariant background gauges.  We find that our 
non-perturbative result is both gauge invariant and gauge parameter $\alpha$ independent.

\end{abstract} 
\maketitle 
%\narrowtext

\newpage 
\section{Introduction}
The chromo-electric flux tube model of particle production \cite{Casher} which is based on earlier work in QED
by Heisenberg and Euler, Schwinger and Weiskopf \cite{Schwinger}, has been a very useful tool for understanding
aspects of particle production in Heavy Ion Colliders such as the RHIC \cite{Baym} at Brookhaven. One of the main 
stumbling blocks for making more realistic models which include back reaction has been
the problem of finding approximations to QCD which are not only gauge invariant but
{\em also} independent of the gauge fixing parameter $\alpha$.   Although the background field gauge leads to an action which is gauge invariant, the action still depends on the gauge parameter $\alpha$.  In perturbation theory it is known that at the one loop and two loop level choosing the background Feynman gauge ($\alpha=1$) simplifies calculations and  leads to the same result as a resummation of the Feynman graphs which give a gauge invariant and $\alpha$ independent result  \cite{Cornwall}. Also some recent results using Pinch techniques \cite{Pap} suggest that the background Feynman gauge result may be the correct 
choice to all orders in perturbation theory in the sense of leading to Schwinger Dyson equations
that are gauge invariant. 

In a recent paper, Nayak and van Nieuwenhuizen \cite{gouranga}
calculated the rate for pair production of gluons and $p_T$ distributions from a constant 
chromo-electric field $E^a$ (a=1,....8)
via the  Schwinger mechanism in the background Feynman-t'Hooft gauge ($\alpha$=1). They found that
the results for the $p_T$ distribution of the gluons produced depends on two Casimir/gauge invariant
quantities $E^aE^a$ and $[d_{abc}E^aE^bE^c]^2$ in SU(3). This dependence on these  two gauge invariant quantities also occurs  in quark-antiquark production as found by Nayak in \cite{nayak1}. In the case of quark-antiquark production
there is no dependence on the gauge parameter and hence the result found in \cite{nayak1} is trivially
gauge invariant and gauge parameter independent. However, in the case of gluon pairs there is a 
gauge parameter dependence coming from the gauge fixing term and hence one needs to show that
the results obtained for case of  gluon pairs for $\alpha$=1 gauge in \cite{gouranga} is 
also valid in any arbitrary covariant gauge parameter $\alpha$.  In this paper we prove that
the non-perturbative result obtained in the paper \cite{gouranga} for the gluon pair case
in $\alpha$=1 gauge, is the correct gauge invariant and gauge parameter
$\alpha$ independent result.
Thus we find that for this non-perturbative process, the background Feynman-t'Hooft gauge {\em does} give the  correct gauge invariant {\em and}  $\alpha$ independent result. 
 Now that we have an 
effective action that leads to gauge invariant results for constant external fields, we can 
use this action  with  arbitrary time (and space) dependent  background fields that obey the 
generalized  Yang-Mills
equations and thus study the back reaction problem in analogy to what was done by Cooper and Mottola for the Electric Field problem \cite{cooper}.  This will be the
subject of a future paper. 

\section{Background Field Method of QCD and Schwinger Mechanism in Pure Gauge
Theory}

In the background field method of QCD the gauge field is the sum of a classical
background field and  the quantum gluon field:
\bea
A_\mu^a ~\rightarrow ~ A_\mu^a~+~Q_\mu^a
\label{aq}
\eea
where in the right hand side $A_\mu^a$ is the classical background field
and $Q_\mu^a$ is the quantum gluon field. The gauge field Lagrangian density is
given 
by 
\bea
{\cal L}_{\rm gauge}~=~-\frac{1}{4} F_{\mu \nu}^a[A+Q] F^{\mu \nu a}[A+Q].
\label{total}
\eea
The background gauge fixing is given by \cite{thooft}
\bea
D_\mu[A]Q^{\mu a}~=~0,
\label{gfix}
\eea
where the covariant derivative is defined by 
\bea
D_\mu^{ab}[A]~=~\delta^{ab} \partial_\mu~+~gf^{abc}A_\mu^c.
\eea
The gauge fixing Lagrangian density is 
\bea
{\cal L}_{\rm gf}~=~-\frac{1}{2\alpha} [D_\mu[A]Q^{\mu a}]^2
\label{gf}
\eea
where $\alpha$ is any arbitrary gauge parameter, and the corresponding ghost contribution is given by

\be
{\cal{L}}_{ghost} ~=~ \overline{\chi}^a D_{\mu}^{ab}[A]
D^{\mu ,bc}[A+Q]\chi^c~=~\overline{\chi}^a~K^{ab}[A,Q]~\chi^b.   
\label{ghos1}
\ee

Now adding eqs. (\ref{total})
and (\ref{gf})  and (\ref{ghos1})we get the Langrangian density for gluons interacting with a
classical background field
\bea
{\cal L}_{\rm gluon}~&&=~-\frac{1}{4} F_{\mu \nu}^a[A+Q] F^{\mu \nu a}[A+Q]
~-~\frac{1}{2\alpha} [D_\mu[A]Q^{\mu a}]^2 \nonumber \\
&&+~\overline{\chi}^a~K^{ab}[A,Q]~\chi^b.  
\label{lgluon}
\eea

To discuss gluon pair production at the one-loop level on considers just the part of this 
Lagrangian which is quadratic in quantum fields.  This quadratic Lagrangian is invariant
under a restricted class of gauge transformations. 
The quadratic Lagrangian  for 
a pair of gluon interacting with background field $A_\mu^a$ is given by
\bea
{\cal L}_{\rm gg}~=~\frac{1}{2}Q^{\mu a} M^{ab}_{\mu \nu}[A] Q^{\nu b}
\label{fulll}
\eea
where 
\bea
M^{ab}_{\mu \nu}[A]~=~
\eta_{\mu \nu} [D_\rho(A)D^\rho(A)]^{ab}~-~2gf^{abc}F_{\mu \nu}^c
~+~(\frac{1}{\alpha}-1)[D_\mu(A) D_\nu(A)]^{ab}
\label{mab}
\eea
with $\eta_{\mu \nu}~=~(-1,+1,+1,+1)$. 

For our purpose we write 
\bea
M^{ab}_{\mu \nu}[A]~=~
{M^{ab }_{\mu \nu,}}_{\alpha=1}[A]~+~\alpha^\prime[D_\mu(A) D_\nu(A)]^{ab}
\label{mab1}
\eea
where $\alpha^\prime~=~(\frac{1}{\alpha}-1)$. The matrix elements for the
gauge parameter $\alpha$=1 is given by
\bea
{M^{ab }_{\mu \nu,}}_{\alpha=1}[A]~= 
\eta_{\mu \nu} [D_\rho(A)D^\rho(A)]^{ab}~-~2gf^{abc}F_{\mu \nu}^c
\label{mabalpha1}
\eea
which was studied in \cite{gouranga}.
In this approximtaion the ghost Lagrangian density (which we will discuss separately) is given by
\be
{\cal{L}}_{ghost} ~=~ \overline{\chi}^a D_{\mu}^{ab}[A]
D^{\mu ,bc}[A]\chi^c~=~\overline{\chi}^a~K^{ab}[A]~\chi^b 
\label{ghos2}
\ee

The vacuum-to-vacuum transition amplitude in pure gauge theory
in the presence of a background field $A_\mu^a$ is given by:
\bea
_+<0|0>^A_- ~=~
\int [dQ] [d\chi] [d\bar{\chi}]~e^{i{(S+S_{gf}+S_{\rm {ghost}})}}.
\eea
For the gluon pair part this can be written by
\bea
_+<0|0>^A_- ~=~\frac{Z[A]}{Z[0]}~=~
\frac{\int [dQ]~e^{i\int d^4x~Q^{\mu a}M_{\mu \nu}^{ab}[A]Q^{\nu b}}}{\int [dQ]~e^{i\int 
d^4x~Q^{\mu a}M_{\mu \nu}^{ab}[0]Q^{\nu b}}}~=~e^{iS^{(1)}_{eff}}
\label{vac1}
\eea
where $S^{(1)}_{eff}$ is the one-loop effective action. The non-perturbative 
real gluon production is related to the
imaginary part of the effective action $S^{(1)}_{eff}$ which is physically due to the
instability of the QCD vacuum in the presence of the background field.
The above equation can be written as
\bea
_+<0|0>^A_- ~=~\frac{Z[A]}{Z[0]}~=~
\frac{{\rm Det^{-1/2}}M_{\mu \nu}^{ab}[A]}{{\rm Det^{-1/2}}M_{\mu \nu}^{ab}[0]}
~=~e^{iS^{(1)}_{eff}}
\label{vac2}
\eea
which gives
\bea
S^{(1)}_{eff}~=~-i {\rm Ln} 
\frac{(\rm Det [M_{\mu \nu}^{ab}[A]])^{-1/2}}{(\rm Det [M_{\mu \nu}^{ab}[A]])^{-1/2}}
~=~\frac{i}{2}Tr[ {\rm Ln} M_{\mu \nu}^{ab}[A] -{\rm Ln} M_{\mu \nu}^{ab}[0]].
\label{vac3}
\eea

The trace Tr contains an integration over $d^4x$ and 
a sum over color and Lorentz indices. 
To the above action, we need to add the ghost action. 
The ghost action is gauge independent and eliminates the unphysical gluon degrees of 
freedom.  The one-loop effective action for the ghost in the background field $A_\mu^a$
is given by
\bea
S^{(1)}_{ghost}~=~-i {\rm Ln}(Det~K)~=~-i~Tr \int_0^\infty~\frac{ds}{s} [
e^{is~[K[0]+i\epsilon]}~ -e^{is~[K[A]+i\epsilon]}]
\eea
where $K^{ab}[A]$ is given by (\ref{ghos2}). 
Since the total action is the sum of the 
gluon and ghost actions, the gauge parameter dependent part proportional 
to $(\frac{1}{\alpha} -1)$ can
be evaluated as an addition to the $\alpha=1$ result and discussed seperately from the 
$\alpha=1$ calculation done earlier. In what follows we will assume when discussing the
$\alpha=1$ calculation that we have included the ghost contribution. 

The non-perturbative gluon pair production per unit volume per unit time is
related to the imaginary part of this effective action via
\bea
\frac{dN}{dtd^3x}~\equiv~{\rm Im} {\cal L}_{eff}~=~\frac{{\rm Im} S^{(1)}_{eff}}{d^4x}.
\label{nongluon}
\eea
This is the general formulation of 
Schwinger mechanism in pure gauge theory where
$M^{ab}_{\mu \nu}[A]$ is given by eq. (\ref{mab1}) and
${M^{ab }_{\mu \nu,}}_{\alpha=1}[A]$ is given by eq. (\ref{mabalpha1}).

\section{Schwinger Mechanism in Pure Gauge Theory in Arbitrary Covariant 
$\alpha$ Gauge}

Using the above formalism the Schwinger mechanism for gluon pair production
was studied in \cite{gouranga} in $\alpha$=1 gauge. In this $\alpha$ =1 gauge 
the final expression for the number of non-perturbative
gluon (pair) production per unit time per unit volume and per unit transverse
momentum from constant chromo-electric field $E^a$ is given by \cite{gouranga}

\bea
\frac{dN_{g,g}}{dt d^3x d^2p_T}~
=~\frac{1}{4\pi^3} ~~ \sum_{j=1}^3 ~
~|g\lambda_j|~{\rm Ln}[1~+~e^{-\frac{ \pi p_T^2}{|g\lambda_j|}}].
\label{1}
\eea
where
\bea
&&~\lambda_1^2~=~\frac{C_1}{2}~[1-{\rm cos}~ \theta],  \nonumber \\
&&~\lambda_2^2~=~\frac{C_1}{2}~[1+{\rm cos}(\frac{\pi}{3}-\theta)],  \nonumber \\
&&~\lambda_3^2~=~\frac{C_1}{2}~[1+{\rm cos}(\frac{\pi}{3}+\theta)],
\label{lm}
\eea
and
\bea
cos^3\theta~=-1+6 C_2/C_1^3.
\label{theta}
\eea
They depend only on the two Casimir/gauge invariants for SU(3)
\bea
C_1~=~E^aE^a, ~~~~~~~~~~~~~~~~~~~~~~~
C_2~=~[d_{abc}E^aE^bE^c]^2,
\label{casm}
\eea
where $a,~b,~c$ ~=~1,...,8 are the color 
indices of the adjoint representation of the gauge group 
SU(3). In this paper we will check if this result remains same for any
arbitrary gauge parameter $\alpha$.

In the arbitrary gauge parameter case, 
we have to evaluate the one-loop effective action (eq. (\ref{vac3})) 
which is given by
\bea
S^{(1)}_{eff}~=
~\frac{i}{2}Tr[ {\rm Ln} M_{\mu \nu}^{ab}[A] -{\rm Ln} M_{\mu \nu}^{ab}[0]],
\label{vacc3}
\eea
where
$M^{ab}_{\mu \nu}[A]$ is given by eq. (\ref{mab1}) and
${M^{ab }_{\mu \nu,}}_{\alpha=1}[A]$ is given by eq. (\ref{mabalpha1}).
To evaluate the trace in eq. (\ref{vacc3}) we can evaluate the trace of
\bea
Tr~Ln[ M^{ab}_{\mu \lambda }[A] \eta^{\lambda \nu}]~=~ Tr~Ln[ 
{{{M^{ab }_{\mu \lambda,}}}}_{\alpha=1}[A]\eta^{\lambda \nu}~+~
\alpha^\prime[D_\mu(A) D^\nu(A)]^{ab}],
\label{trmab}
\eea
where we added $\eta^{\lambda \nu}$ inside
the trace in $Tr~Ln[ M^{ab}_{\mu \nu }[A]]$
because this cancels against the free part
$Tr~Ln[ M^{ab}_{\mu \nu }[0]]$ in eq. (\ref{vacc3}).
The above equation can be split into the $\alpha$=1 part (which was
studied in (\cite{gouranga})) and a gauge parameter $\alpha$ dependent part
as follows
\bea
&& Tr~Ln[ M^{ab}_{\mu \lambda }[A] \eta^{\lambda \nu}]~=~ Tr~Ln[ 
{{M^{ab }_{\mu \lambda,}}}_{\alpha=1}[A] \eta^{\lambda \nu}]~\nonumber \\
&& +~ Tr~Ln[ \delta^{ab} \delta_{\mu}^{\nu}~+~ \alpha^\prime
{{{M^{-1}}^{ac }_{\mu}}^\lambda,}_{\alpha=1}[A] ~[D_\lambda(A) D^\nu(A)]^{cb}].~
\label{trmab1}
\eea

We follow the $\alpha$=1 case \cite{gouranga} and
assume that the constant chromo-electric field is along 
the z-axis (the beam direction) and we choose the gauge $A_0^a$= 0
so that $A_3^a=-E^a{\hat x}^0$. The color indices (a=1,....8) are arbitrary.
Now using the relation 
\be
[D_\mu, D_\nu]^{ab} = -  gf^{abc} F_{\mu \nu}^c
\ee
we find that
\bea
[D_\nu (A)M^{\nu \mu}_{\alpha=1}[A]]^{ab}~=~[D^2(A) D^\mu(A)]^{ab}
\label{dmumab}
\eea
and
\bea
[M^{\mu \nu}_{\alpha=1}[A]D_{\nu}(A)]^{ab}~=~[D^\mu(A)D^2(A) ]^{ab}
\label{mabdmu}
\eea
where 
${M^{ab}_{\mu \nu,}}_{\alpha=1}[A]$ is given by eq. (\ref{mabalpha1}).
We can use these identities to rewrite Eq. (\ref{trmab1})
in a very convenient fashion
which will allow us to show the $\alpha$ independence of the result. 
First we multiply ${{M^{-1}}^{ab}_{\mu \nu,}}_{\alpha=1}[A]$ from the 
left in eq. (\ref{mabdmu}) and obtain
\bea
[ {{M^{-1}}_{\mu \lambda,}}_{\alpha=1}[A]
{M^{\lambda \nu}}_{\alpha=1}[A]
D_{\nu}(A)]^{ab}~=[~
{{M^{-1}}_{\mu \lambda,}}_{\alpha=1}[A]
D^\lambda (A)D^2(A) ]^{ab}
\eea
which gives
\bea
D_{\mu}^{ab}(A)~=[~{{M^{-1}}_{\mu \lambda,}}_{\alpha=1}[A] D^\lambda (A)D^2(A) ]^{ab}
\label{minvmdmu}
\eea
Now multiplying $[\frac{1}{D^2(A)}]^{ab}$ from right in the above equation we get
\bea
[D_{\mu}(A)\frac{1}{D^2(A)}]^{ab}~=[~{{M^{-1}}_{\mu \lambda,}}_{\alpha=1}[A] D^\lambda (A)]^{ab}.
\label{dmubyd2}
\eea
Multiplying $D_{\nu}^{ab}(A)$ from the right in the above equation we get
\bea
[D_{\mu}(A)\frac{1}{D^2(A)} D^{\nu}(A) ]^{ab}
~=[~{{M^{-1}}_{\mu \lambda,}}_{\alpha=1}[A] D^\lambda (A) D^{\nu}(A) ]^{ab}.
\label{dmubyd2dnu}
\eea
Using eq. (\ref{dmubyd2dnu}) in eq. (\ref{trmab1}) we get

\bea
&& Tr~Ln[ {M^{ab}_{\mu}}^{\nu}[A] ]~=~ Tr~Ln[ 
{{{M^{ab }_{\mu }}^{\nu,}}}_{,~\alpha=1}[A] ]~\nonumber \\
&& +~ Tr~Ln[ \delta^{ab} \delta_{\mu}^{\nu}~+~ \alpha^\prime
[D_{\mu}(A)\frac{1}{D^2(A)} D^{\nu}(A) ]^{ab}]
\label{trmab2}
\eea

Now let us evaluate the trace of the last term
\bea
&& Tr~Ln[ \delta^{ab} \delta_{\mu}^{\nu}~+~ \alpha^\prime [D_{\mu}(A)\frac{1}{D^2(A)} D^{\nu}(A) ]^{ab}] ~=
~Tr~[ 
\alpha^\prime [D_{\mu}(A)\frac{1}{D^2(A)} D^{\nu}(A) ]^{ab}] \nonumber \\
&&~-~Tr~[ 
\frac{{\alpha^\prime}^2}{2}~ [
D_{\mu}(A)
\frac{1}{D^2(A)} D^{\nu}(A)
 ]^{ab}]
~+~Tr~[ 
\frac{{\alpha^\prime}^3}{3}~ [
D_{\mu}(A) \frac{1}{D^2(A)} D^{\nu}(A)
 ]^{ab}] \nonumber \\
&&
~-~Tr~[ 
 \frac{{\alpha^\prime}^4}{4}~ [
D_{\mu}(A)
\frac{1}{D^2(A)} D^{\nu}(A)
 ]^{ab} ]
~+~Tr~[ 
\frac{{\alpha^\prime}^5}{5}~ [
D_{\mu}(A)
\frac{1}{D^2(A)} D^{\nu}(A) ]^{ab} ]
~+~.........
\label{trmab6}
\eea
Summing the series we obtain
\be  {\rm Ln} (1+\alpha^\prime) Tr~ [
D_{\mu}(A)
\frac{1}{D^2(A)} D^{\nu}(A)
 ]^{ab} ]  \label{trans}
 \ee
The trace $Tr$ equals to
\bea
Tr=tr~tr_{ab}~tr_{\mu \nu}
\eea
where 
\bea
tr [{\cal O}]~=~\int d^4x <x|{\cal O}|x>.
\eea
Using the cyclic properties of the full trace (Tr) we then obtain:
\bea
Tr[ D_{\mu}(A) \frac{1}{D^2(A)} D^{\nu}(A) ]^{ab} ]~=Tr[D^{\nu}(A)D_{\mu}(A) \frac{1}{D^2(A)}  ]^{ab} ]=~8~\int d^4x. 
\eea
Using this in eq. (\ref{trans}) we find
\bea
&& Tr~Ln[ \delta^{ab} \delta_{\mu}^{\nu}
~+~ \alpha^\prime [D_{\mu}(A)\frac{1}{D^2(A)} D^{\nu}(A) ]^{ab}] ~\nonumber \\
&&~=~8 ~Ln[1+\alpha^\prime]~\int d^4x ~ =~-8 ~Ln[\alpha]~\int d^4x 
\label{trmab7}
\eea

In case of free gluons ($M_{\mu \nu }^{ab}[0]$) we get:
\bea
Tr~Ln[ \delta^{ab} \delta_{\mu}^{\nu}
~+~ \alpha^\prime [\partial_{\mu}\frac{1}{\partial^2} \partial^{\nu}\delta^{ab}]] ~
~=~8 ~Ln[1+\alpha^\prime]~\int d^4x 
~=~-8 ~Ln[\alpha]~\int d^4x 
\label{trmab8}
\eea
Hence from eqs. (\ref{vacc3}), (\ref{trmab2}), (\ref{trmab7}) and (\ref{trmab8})
we get:
\bea
&& S^{(1)}_{eff}~=~\frac{i}{2}Tr[Ln {M^{ab}_{\mu}}^{\nu}[A] ~-~ 
Ln {M^{ab}_{\mu}}^{\nu}[0] ]~=~ \nonumber \\
&& \frac{i}{2}Tr[Ln {{{M^{ab }_{\mu }}^{\nu,}}}_{\alpha=1}[A] ~
-~Ln {{{M^{ab }_{\mu }}^{\nu,}}}_{\alpha=1}[0] ]~=
{S^{(1)}_{eff,}}_{\alpha=1}. 
\label{trfin}
\eea
Hence the gauge parameter dependence on $\alpha$  exactly cancels from the interacting part 
and the  free part  and we get a gauge parameter independent (and gauge invariant!) 
result for gluon production which is the  same as that obtained in the
$\alpha$=1 gauge. 

\section{Conclusions}
We have studied non-perturbative gluon (pair) production from a constant chromo-electric field with arbitrary color via the  Schwinger mechanism 
in a class of covariant background gauges described by the gauge parameter $\alpha$ by directly evaluating the path 
integral. We find that the non-perturbative gluon production rate and its 
$p_T$ distribution are independent of the gauge parameter $\alpha$ and hence
the result is  both gauge invariant and gauge parameter $\alpha$ independent. This result will allow
us to use the effective two gluon action in the Feynman-t'Hooft gauge as a starting point
for gluon pair production and back reaction. 

\acknowledgments

This work was supported 
in part by the National Science Foundation, grants PHY-0071027, 
PHY-0098527, PHY-0354776 and PHY-0345822.
The authors would like to thank
the Santa Fe Institute for its hospitality during the course of this work.

\end{document}